\documentclass[conference]{IEEEtran}

\IEEEoverridecommandlockouts
\usepackage{graphicx}
\graphicspath{ {figures/} }
\usepackage{array}

\usepackage{url}
\usepackage{amsmath}
\usepackage{float}
\usepackage{pgfplots}
\usepackage{color,soul}
\usepackage{hyperref}

\definecolor{lightgreen}{RGB}{225,250,222}
\pgfplotsset{width=10cm,compat=1.9}
\usepackage{cite}
\usepackage{amsmath,amssymb,amsfonts}
\usepackage{algorithmic}
\usepackage{graphicx}
\usepackage{textcomp}
\usepackage{xcolor}
\usepackage{float}
\usepackage{xcolor}
\usepackage{tikz}
\usepackage{adjustbox}
\usepackage[figurename=Figure]{caption}
\def\BibTeX{{\rm B\kern-.05em{\sc i\kern-.025em b}\kern-.08em
    T\kern-.1667em\lower.7ex\hbox{E}\kern-.125emX}}
\begin{document}

\title{End-to-End Bangla Speech Synthesis\\}

%
\definecolor{bblue}{HTML}{4F81BD}
\definecolor{rred}{HTML}{C0504D}
\definecolor{ggreen}{HTML}{9BBB59}
\definecolor{ppurple}{HTML}{9F4C7C}


\author{\IEEEauthorblockN{Prithwiraj Bhattacharjee, Rajan Saha Raju, Arif Ahmad, M. Shahidur Rahman}
\IEEEauthorblockA{Department of Computer Science and Engineering, 
Shahjalal University of Science \& Technology, Sylhet, Bangladesh\\
Email: \{prithwiraj12, rajan10\}@student.sust.edu, \{arif\_ahmad-cse, rahmanms\}@sust.edu}
}

\maketitle

\begin{abstract}
Text-to-Speech (TTS) system is a system where speech is synthesized from a given text following any particular approach. Concatenative synthesis, Hidden Markov Model (HMM) based synthesis, Deep Learning (DL) based synthesis with multiple building blocks, etc. are the main approaches for implementing a TTS system. Here, we are presenting our deep learning-based end-to-end Bangla speech synthesis system. It has been implemented with minimal human annotation using only 3 major components (Encoder, Decoder, Post-processing net including waveform synthesis). It does not require any frontend preprocessor and Grapheme-to-Phoneme (G2P) converter. Our model has been trained with phonetically balanced 20 hours of single speaker speech data. It has obtained a 3.79 Mean Opinion Score (MOS) on a scale of 5.0 as subjective evaluation and a 0.77 Perceptual Evaluation of Speech Quality(PESQ) score on a scale of [-0.5, 4.5] as objective evaluation. It is outperforming all existing non-commercial state-of-the-art Bangla TTS systems based on naturalness.
\end{abstract}


\section{Introduction}
Text-to-Speech (TTS) system is a system that generates speech from any given text. Both Digital Signal Processing (DSP) and Natural Language Processing( NLP) are being used to develop such a system. In previous years, some great works have been done on TTS research. But many things should be taken under consideration for developing a better TTS system. If we preciously indicate language-specific TTS systems then there lies a vast area to work on. Bangla language is the fifth-most spoken language in the world. Technology like Bangla TTS system has a great impact on our daily life. But no significant implementation of Bangla TTS is available because all the existing systems are based on traditional approaches. So, we planned to take an initiative to develop a better Bangla TTS system by using the current state-of-the-art technologies.
\begin{figure}[H]
  \begin{center}
     \includegraphics[width=8.5cm, height=3.5cm]{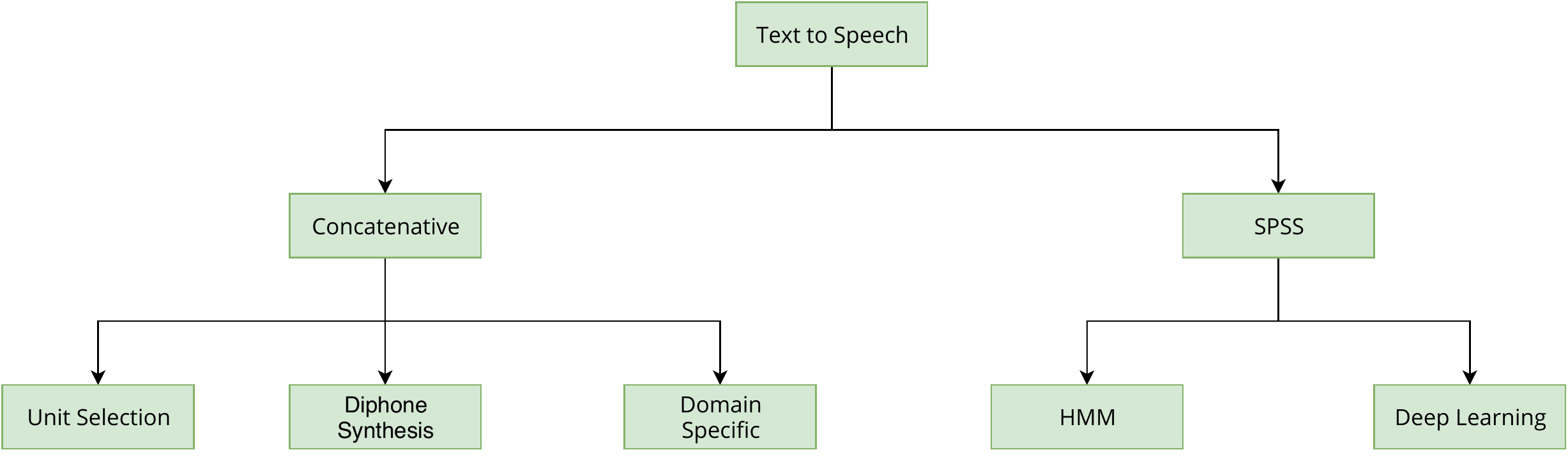}
     \caption{Classification of TTS system}{}
      \label{tts}
  \end{center}
\end{figure}
TTS systems are mainly classified into two major parts. \textbf{Concatenative TTS} and \textbf{Statistical Parametric Speech Synthesis (SPSS) TTS}. Some concatenative TTS systems are unit selection, diphone synthesis, and domain-specific synthesis. Unit selection requires pre-recorded speech and diphone synthesis needs all the sound-to-sound transitions or diphones. Hidden Markov Model (HMM) based TTS and Deep Neural Network (DNN) based TTS are SPSS TTS. Major components of traditional DNN based TTS systems are frontend preprocessing, acoustic modeling, duration modeling, and vocoder. In recent times, end-to-end TTS systems have become very popular. This neural network-based TTS system neither needs any pre-recorded speech nor any building blocks such as frontend preprocessing. Though researches on Bangla TTS started a long time back, we continued the legacy by doing researches on the latest TTS technologies. TTS system is very effective for physically impaired or blind people. Again modern business, airports, and all other notable sectors use the TTS system. These are some of our research motivations. Our discriminative research contributions are as follows:
\begin{itemize}
    \item This is the first-ever Bangla end-to-end TTS system. It does not require complex linguistic and acoustic features as input.
    \item We built phonetically rich studio-quality speech data containing more than 40 hours of speech which more than any other publicly available dataset with the help of two professional voice artists.
    \item In terms of naturalness, the system outperforms among all other non-commercial Bangla TTS systems.
\end{itemize}

\section{Related Work}
The very first attempt of implementing Bangla TTS system was concatenative developed by Alam et. al. \cite{alam2007text}. It was a unit selection-based TTS system named Katha where they used festival\footnote{http://www.cstr.ed.ac.uk/projects/festival/} toolkit for front-end preprocessing. In 2009, a diphone concatenation-based TTS system Subachan was developed by Naser et. al. \cite{naser2010implementation}. In 2014, Mukharjee et. al. \cite{mukherjee2014bengali} developed the first HMM-based SPSS system for Bangla. But over smoothness of acoustic modeling and limitations of vocoders resulted in producing less natural output in HMM-based system. In 1993, Weijters et. al. \cite{weijters1993speech} first gave an idea of implementing neural networks for acoustic modeling. But when Hinton et. al. \cite{hinton2012deep}, ze et. al. \cite{ze2013statistical} and Ling et. al. \cite{ling2015deep} individually implemented Deep neural networks (DNNs) for acoustic modeling, a great enhancement occurred in TTS research. Recurrent Neural Networks (RNNs) of Ze et. al. \cite{ze2013statistical}, Long Short-Term Memory RNNs (LSTM-RNNs) of Qian et. al. \cite{qian2014training}, and deep Bidirectional LSTM-RNNs (BLSTM-RNNs) of Fan et. al. \cite{fan2014tts} significantly improved the performance of SPSS systems. Gutkin et. al. \cite{gutkin2016tts} from google released a commercial Bangla TTS system but the project is not publicly available. Although Google has released some language-specific resources on github\footnote{https://github.com/google/language-resources/} for enhancing Bangla TTS research. Potard et. al. \cite{potard2016idlak} developed an open-source TTS system named Idlak Tangle taking help from another open-source speech recognition system named Kaldi proposed by Povey et. al. \cite{povey2011kaldi}. Wu et. al. \cite{wu2016merlin} from Edinburgh University developed another DNN based open source TTS system named Merlin where they used ossian\footnote{https://github.com/CSTR-Edinburgh/Ossian} and festival frontend preprocessors and WORLD vocoder of Morise et. al.\cite{morise2016world} for synthesizing speech. Taking help from these open source tools, Deka et. al. \cite{deka2019development} developed a TTS system for a low resourced language. In 2019, Raju et. al. \cite{raju2019bangla} also developed a DNN based Bangla SPSS TTS using this Merlin toolkit. In most cases, SPSS methods are better than concatenative approaches but they need a huge amount of human annotation such as language-specific text processing. So, researchers are now interested to develop end-to-end TTS systems where speech is directly generated from (text, speech) pairs without having any preprocessing. Inspiring from this concept, Arik et. al. \cite{arik2017deep}, Gibiansky et. al. \cite{gibiansky2017deep} and Ping et. al. \cite{ping2017deep} from Baidu developed Deep Voice 1, 2, and 3 respectively. Then Sotelo et. al.\cite{sotelo2017char2wav} from MILA developed Char2Wav and Ren et. al. \cite{ren2019fastspeech} from  Microsoft developed Fastspeech. Google has also published some TTS research-based papers named WaveNet by Oord et. al. \cite{oord2016wavenet}, Tacotron by  Wang et. al. \cite{wang2017tacotron} and Tacotron 2 by Shen et. al. \cite{shen2018natural}. Though commercial companies do not release their project publicly, individual researchers release their implementation online. By observing all these cases, we adopted a good implementation released by Keithito\footnote{https://github.com/keithito/tacotron}. It is based on tacotron developed by Wang et. al. \cite{wang2017tacotron}. Thus we implemented our end-to-end Bangla TTS system.

\section{Dataset Collection and Preprocessing}
To build a good TTS system we need to collect significant speech data with corresponding transcriptions. Google has released 3 hours of Bangla speech data. Alam et. al. \cite{alam2010development} from Brac University also released 13 hours of speech data. These are some notable publicly available datasets. But modern research stated that for developing a good TTS system we need at least 20 hours of speech data. So, we started to prepare a standard and notable dataset from scratch in our institution. After the preparation, we did some preprocessing to make it compatible with training the model.  

\subsection{Text Corpus}
From Honnet et.al. \cite{honnet2017siwis}, Sonobe et. al. \cite{sonobe2017jsut}, and Gabdrakhmanov et. al. \cite{gabdrakhmanov2019ruslan} we got the concept of developing a phonetically balanced text corpus. We then collected our text data from various domains confirming that our corpus has all possible punctuations for Bangla. Our final dataset contains more than 12500 utterances. A preview of our dataset is presented in table \ref{speechDataSummary}.

\begin{table}[H]
    \centering
    \begin{tabular}{|l|r|}
    \hline
    Total sentences                   & $12,537$         \\
    \hline
    Total words                       & $1,22,627$       \\
    \hline
    Total unique words                & $24,582$         \\
    \hline
    Minimum words in a sentence       & $3$              \\
    \hline 
    Maximum words in a sentence       & $20$             \\
    \hline 
    Average words in a sentence       & $9.78$           \\
    \hline 
    Total duration of speech (hours)          & $20:14:21$  \\
    \hline
    Average duration of each sentence (seconds)  & $5.81$ \\
    \hline

    \end{tabular}
    \caption{Summary of Dataset}
    \label{speechDataSummary}
\end{table}

\subsection{Speech Data}

After completion of preparing the text corpus, we then started recording the speech data. In our university, a soundproof lab has been built for recording the speech. Two professional voice artists (one male and one female) were employed to record speech for several consecutive hours. The sample rate of wave files is 48KHz. The duration of the collected speech is around 20 hours for each speaker. To make the data compatible with the tacotron model, we did some preprocessing. We removed all the sentences containing less than or equal to 3 words, greater than or equal to 12 words, and their corresponding speech. Then we removed the silence from starting and ending part from all the remaining waveforms. Thus we made our dataset compatible for training the model.


\subsection{Text Normalization}
Text normalization is the process of converting a raw text to its pronounceable form. To remove the ambiguities, nonstandard words should be converted into their standard pronunciation. Table \ref{tab:gt} shows different examples of Bangla text normalization. So, we handled the issues like ambiguities of numerical words, abbreviations, etc., and normalized our full training data for generating more accurate pronunciation.
\begin{table}[ht]
\begin{tabular}{c}
\includegraphics[width=9cm]{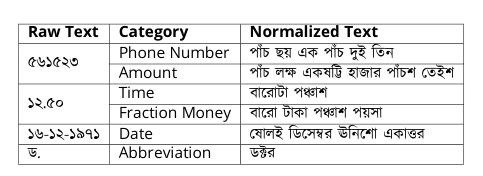}
\end{tabular}
\caption{Bangla Text Normalization}
\label{tab:gt}
\end{table}

\section{System Architecture}
Wang et. al. \cite{wang2017tacotron} proposed that tacotron is an end-to-end generative TTS model based on sequence-to-sequence model by Sutskever et. al. \cite{sutskever2014sequence}. It also uses the attention mechanism of Bahdanau et. al. \cite{bahdanau2014neural}. Tacotron uses Griffin-Lim reconstruction algorithm to generate waveform from spectrogram where Tacotron 2 \cite{shen2018natural} uses WaveNet vocoder to invert mel spectrogram into waveform. But WaveNet vocoder approach is slower than Griffin-Lim based synthesizer. So, we pick this model to develop our end-to-end Bangla TTS system. This system takes characters as input and generates a spectrogram as output. Tacotron TTS system does not need any frontend preprocessing and Grapheme-to-Phoneme (G2P) model. Training starts from scratch with random initialization. The system has three major components which drive the core methodology of the system. The components are given below:
\begin{itemize}
   \item  An Encoder
   \item  An Attention-based Decoder
   \item  A Post-processing Net
\end{itemize}
There are some other components like the CBHG module, Griffin-lim reconstruction, and pre-net. We will now describe these components.
\subsection{CBHG Module}

 \begin{figure}[H]
  \begin{center}
     \includegraphics[width=9cm,height=7cm]{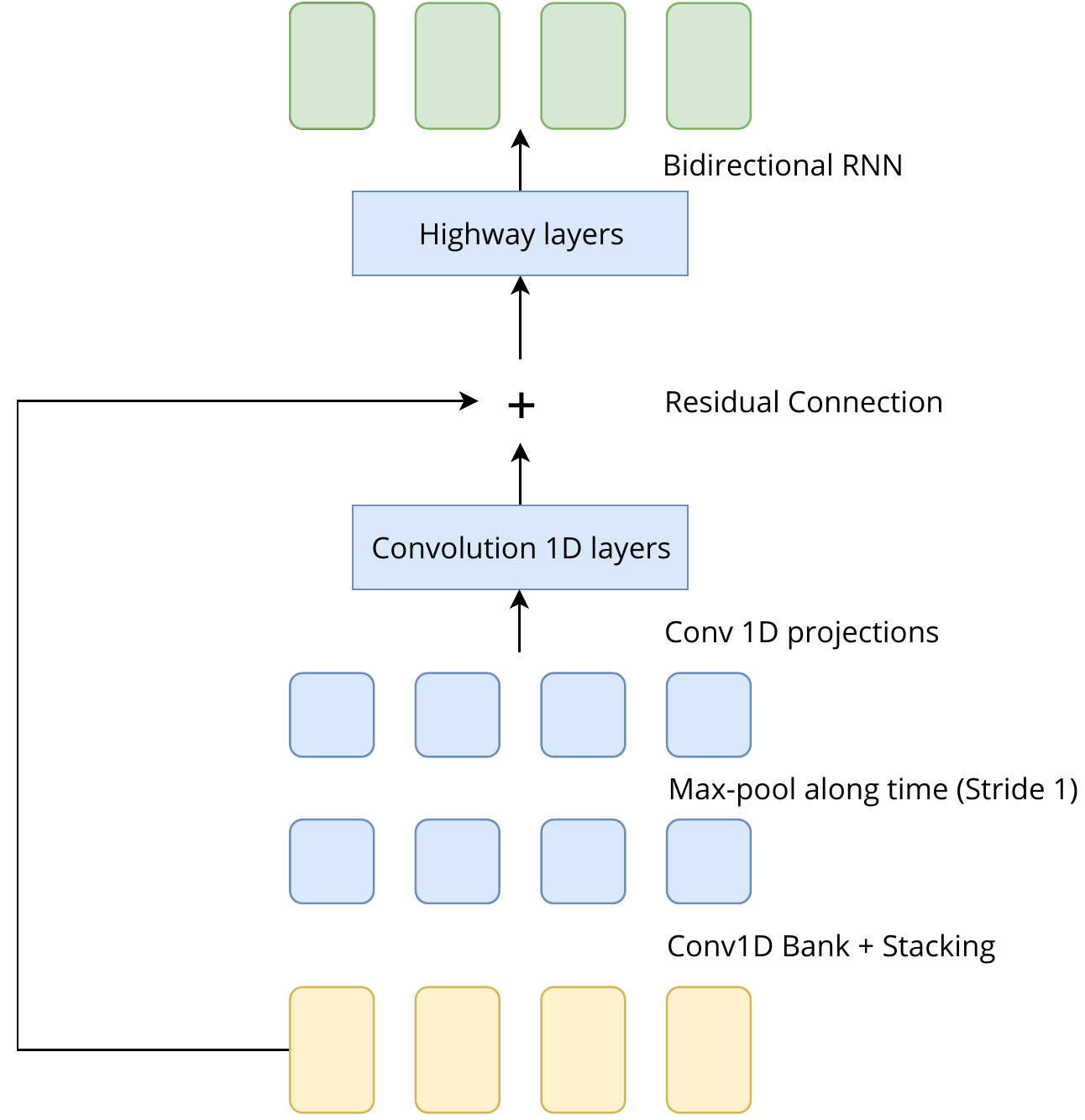}
     \caption{CBHG Module from {\cite[p.~3]{wang2017tacotron}}
}{}
      \label{cbhg}
  \end{center}
\end{figure}
CBHG stands for a 1-D Convolutional Bank, Highway Networks, and a Bidirectional Gated Recurrent Unit (GRU). In CBHG, the input sequence has been passed through 1-D convolution filters. After doing some processing the convoluted output is then added with the original input sequence by residual connections introduced by He et. al. \cite{he2016deep}. Then the output is feed into a multi-layer highway network which is used for extracting high-level features. Lastly, a Bidirectional GRU is used at the top level for extracting sequential features. By using both forward and backward context the GRU completes the extraction. Both batch normalization of Ioffe et. al. \cite{ioffe2015batch} and residual connections etc. improved the generalization. Figure \ref{cbhg} shows the CBHG module.


\subsection{Major Components}
\begin{figure}[H]
  \begin{center}
     \includegraphics[width=9cm,height=9cm]{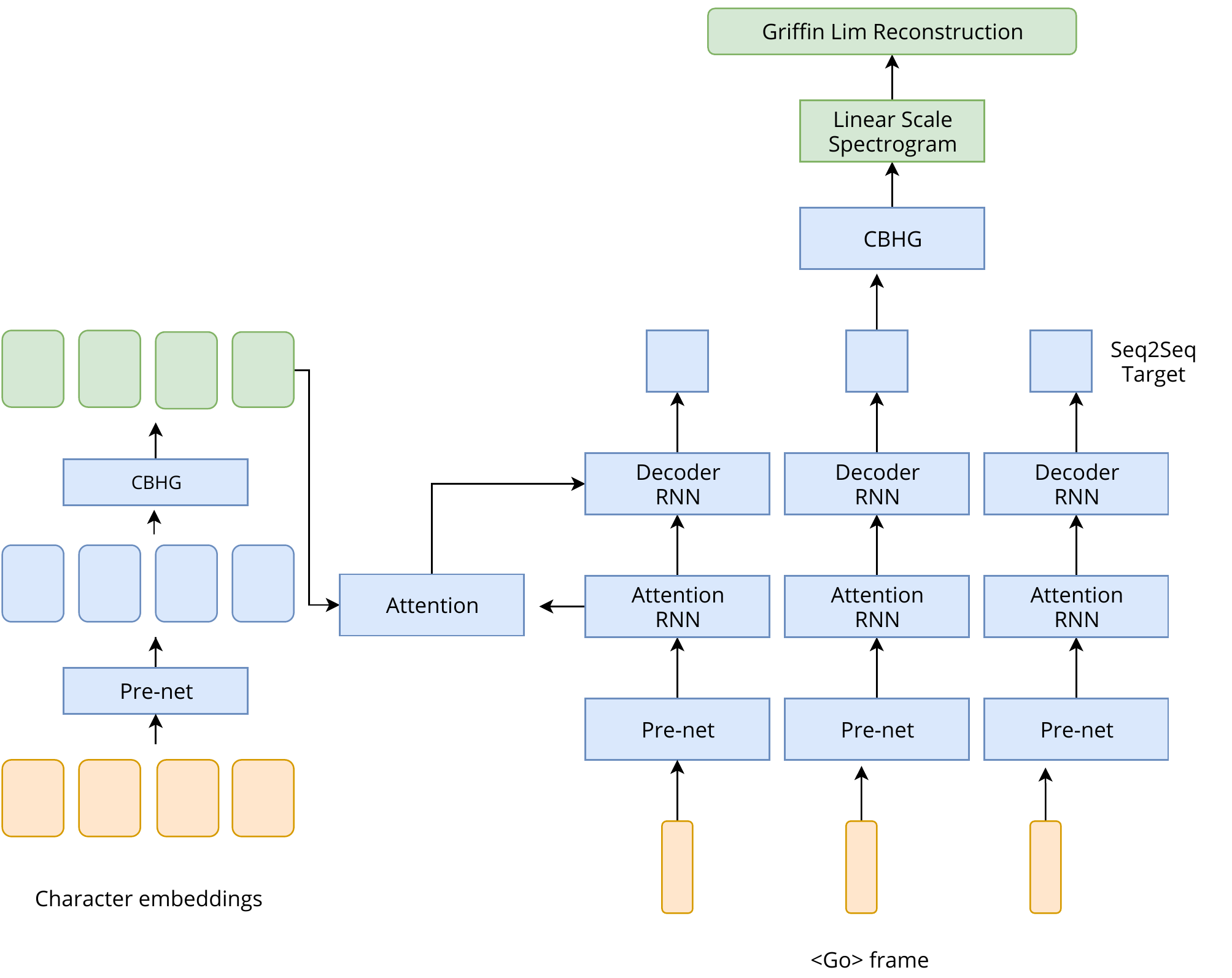}
     \caption{System Architecture from {\cite[p.~2]{wang2017tacotron}}}
     \label{architecture}
  \end{center}
\end{figure}

Tacotron has three major building blocks. An encoder, an attention-based decoder, and a post-processing net. The complete system architecture is shown in figure \ref{architecture}. 
\subsubsection{Encoder}
Sequential representations of text are extracted by an encoder. Character sequences are the inputs for the encoder represented as a one-hot vector. These one-hot vectors are then embedded into a continuous vector representation. For all this to happen, some other components are also being used in the encoder part.
\begin{itemize}
    \item Pre-net
    \item CBHG Module
\end{itemize}
Pre-net is a combination of some non-linear transformations and applied to each embedding. It helps to improve the generalization as well as the convergence of the system. CBHG module transforms the output to a final representation and an attention module uses the modified output for further processing. CBHG-based encoder reduces the overfitting and mispronunciation of words.  

\subsubsection{Attention Based Decoder}
Vinyals et. al.\cite{vinyals2015grammar} attention based decoder is used in our system. The decoder part includes the following things:
\begin{itemize}
    \item Attention RNN
    \item Pre-net
    \item Decoder RNN
    \item Sequence-to-Sequence Target
\end{itemize}
At first, the encoder output is concatenated with the attention RNN and fed into the decoder RNN. An initial pre-net is also fed into the decoder with all zero frames. Decoder consists of a stack of GRUs connected via residual connections. The sequence-to-sequence target model receives the output of the decoder RNN. Then a fixed group of sequence-to-sequence targets started feeding into a CBHG module in the post-processing net part.

\subsubsection{Post Processing Net}
Speech is synthesized in the post-processing net. The post-processing net knows how to predict spectral magnitude based on a linear-frequency scale. It has both forward and backward information and can observe the decoded sequence of target output. Two main sub-components play a vital role in the post-processing net.
\begin{itemize}
    \item CBHG Module
    \item Griffin-Lim algorithm
\end{itemize}
CBHG module takes the sequence-to-sequence target output as input and produces the linear-scale spectrogram. Griffin-Lim algorithm, introduced by Griffin et. al. \cite{griffin1984signal} is used for generating speech from the spectrogram. This is called spectrogram inversion.

\subsubsection{Griffin-Lim Reconstruction}
This algorithm is by far the fastest spectrogram inversion algorithm. It generates speech from the raw spectrogram. Abadi et. al. \cite{abadi2016tensorflow} implemented the Griffin-Lim algorithm with TensorFlow and we followed this procedure. Though this algorithm converges after 50 iterations, we went up to 60 iterations for more accuracy.  

\section{Our System}
\begin{figure}[H]
  \begin{center}
     \includegraphics[width=9cm,height=7cm]{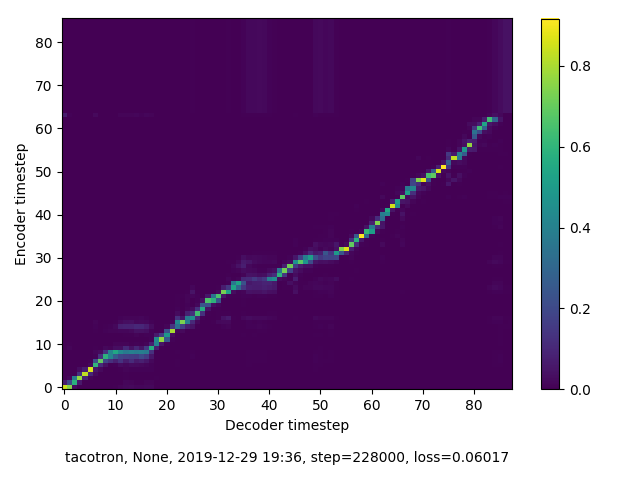}
     \caption{Attention alignments on a test phrase after 228k steps}{}
      \label{graph3}
  \end{center}
\end{figure}

The system has been trained for several days and generates a new model checkpoint after every 1000 iterations. Each checkpoint synthesizes a waveform and generates a graph of attention alignments between encoder and decoder timestamp with loss information. Figure \ref{graph3} shows the attention alignment graph of our model. The best checkpoint has been found after 228k steps of training with a loss of 6.017\%. 
The Y-axis of this graph denotes the encoder timestamp which at each step took an input character as well as its current state and gave an output of 100 real vectors representing the status of each moment. The X-axis denotes the decoder timestamp which reads all the status vectors of the encoder and step-by-step generated audio frames or mel-spectrogram. The diagonal line represents the result of those audio frames based on each input character. The more the graph is diagonally aligned the more the model performs well. 

\section{Results}
There are mainly two types of evaluation that exist to measure the performance of a TTS system. \textbf{Objective Evaluation} and \textbf{Subjective Evaluation}. From these evaluations, we will now show the numerical results as well as the graphical comparisons of different existing Bangla TTS systems.

\subsection{Objective Evaluation}
Objective evaluation is nothing but a mathematical comparison of the generated signal and the original signal. Perceptual Evaluation of Speech Quality (PESQ) \cite{rix2001perceptual} is one of the most adopted metrics for objective evaluation.

\subsubsection{PESQ}
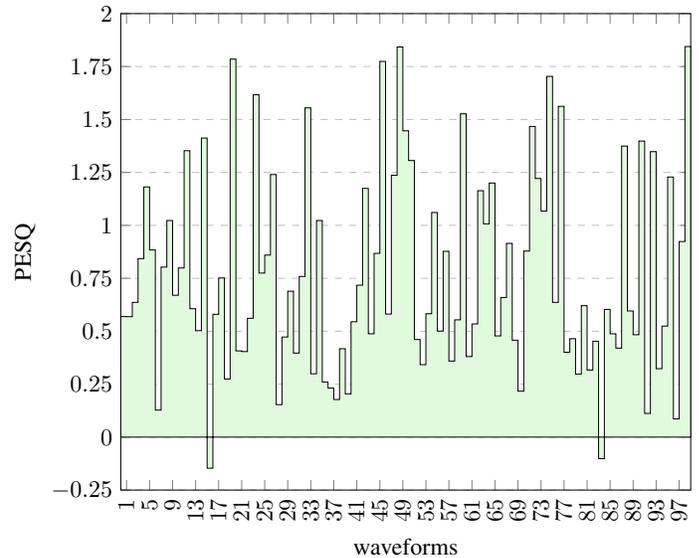
\begin{figure}[h]

\begin{tikzpicture}[thick, scale = .9]
\begin{axis} [
    xlabel = waveforms,
    ylabel = PESQ,
    xmin=0, xmax=99,
    ymin=-0.25, ymax=2.0,
    xtick={1, 5, 9, 13, 17, 21, 25, 29, 33, 37, 41, 45, 49, 53, 57, 61, 65, 69, 73, 77, 81, 85, 89, 93, 97},
    ytick={-0.25,0.0,0.25,0.5,0.75,1.0,1.25,1.5,1.75,2.0},
    x tick label style={rotate=90},
    ymajorgrids=true,
    grid style=dashed,
]
\addplot[
    const plot,
    fill=lightgreen,
    draw=black
]
coordinates {
(0, 0.5694905519485474)
(1, 0.568580150604248)
(2, 0.6362547874450684)
(3, 0.8428533673286438)
(4, 1.1815284490585327)
(5, 0.8837481141090393)
(6, 0.12757563591003418)
(7, 0.8029289245605469)
(8, 1.022782564163208)
(9, 0.6699384450912476)
(10, 0.7986712455749512)
(11, 1.352550983428955)
(12, 0.6064537763595581)
(13, 0.5027586221694946)
(14, 1.412122368812561)
(15, -0.1468675559386611)
(16, 0.5798889398574829)
(17, 0.7515757083892822)
(18, 0.2745457887649536)
(19, 1.785064697265625)
(20, 0.4072309732437134)
(21, 0.40398523211479187)
(22, 0.5612151026725769)
(23, 1.616726279258728)
(24, 0.774720311164856)
(25, 0.8595393896102905)
(26, 1.2400678396224976)
(27, 0.1532829850912094)
(28, 0.47274091839790344)
(29, 0.6891055703163147)
(30, 0.3968825936317444)
(31, 0.7587707042694092)
(32, 1.554849624633789)
(33, 0.2990737855434418)
(34, 1.0229475498199463)
(35, 0.260097861289978)
(36, 0.2318117767572403)
(37, 0.1774924397468567)
(38, 0.4174339175224304)
(39, 0.20437590777873993)
(40, 0.5448636412620544)
(41, 0.7173309326171875)
(42, 1.1745216846466064)
(43, 0.4880587160587311)
(44, 0.8676734566688538)
(45, 1.7742983102798462)
(46, 0.5810681581497192)
(47, 1.236672282218933)
(48, 1.8423633575439453)
(49, 1.4465290307998657)
(50, 1.3062143325805664)
(51, 0.4609638750553131)
(52, 0.34224995970726013)
(53, 0.582627534866333)
(54, 1.0607492923736572)
(55, 0.5005271434783936)
(56, 0.8774361610412598)
(57, 0.3591432571411133)
(58, 0.5538172125816345)
(59, 1.5266411304473877)
(60, 0.38048088550567627)
(61, 0.5345335602760315)
(62, 1.1637123823165894)
(63, 1.0071808099746704)
(64, 1.1999424695968628)
(65, 0.4779213070869446)
(66, 0.6592240929603577)
(67, 0.9149394035339355)
(68, 0.45765843987464905)
(69, 0.21686294674873352)
(70, 0.8785908818244934)
(71, 1.467122197151184)
(72, 1.2214059829711914)
(73, 1.0677952766418457)
(74, 1.7034775018692017)
(75, 0.6362763047218323)
(76, 1.5619698762893677)
(77, 0.40077680349349976)
(78, 0.46480298042297363)
(79, 0.29778990149497986)
(80, 0.6206737756729126)
(81, 0.3167427182197571)
(82, 0.4528432786464691)
(83, -0.1014455184340477)
(84, 0.6030200719833374)
(85, 0.48800894618034363)
(86, 0.42032715678215027)
(87, 1.3744852542877197)
(88, 0.5957123637199402)
(89, 0.4835098683834076)
(90, 1.3981225490570068)
(91, 0.11184773594141006)
(92, 1.348119854927063)
(93, 0.32310065627098083)
(94, 0.5242664217948914)
(95, 1.2282500267028809)
(96, 0.08612050861120224)
(97, 0.9236672282218933)
(98, 1.844363355439453)
(99, 1.4155290307998657)
}
\closedcycle;
\label{stimuli_pesq}
\end{axis}
\sloppy
\end{tikzpicture}
\caption{PESQ score of 100 synthesized wave}
\label{fig:pesq_stimuli}
\end{figure}

There are different types of PESQ measurement exist. MOS-LQO and raw-PESQ are mostly being used. The ranges of raw-PESQ and MOS-LQO are [-0.5,4.5] and [1.0, 5.0] respectively. We measured the raw-PESQ score for evaluating our model performance. As both the original and generated waveforms are required to calculate PESQ, we took 100 random sentences and their original waveforms from the test dataset. From those 100 sentences, we synthesized the waveforms using our TTS system. After that, the original and generated waveforms of corresponding sentences were sent to the PESQ system simultaneously. Figure \ref{fig:pesq_stimuli} shows the PESQ score of those 100 generated waveforms. We then calculated the average PESQ score of those 100 waveforms and got \textbf{0.77} as the average PESQ score.

\subsection{Subjective Evaluation}
Subjective evaluation is the user-oriented evaluation. Mean Opinion Score (MOS) \cite{viswanathan2005measuring} is one of the most popular metrics for subjective evaluation and we also calculated the MOS to evaluate our TTS system.

\subsubsection{MOS}
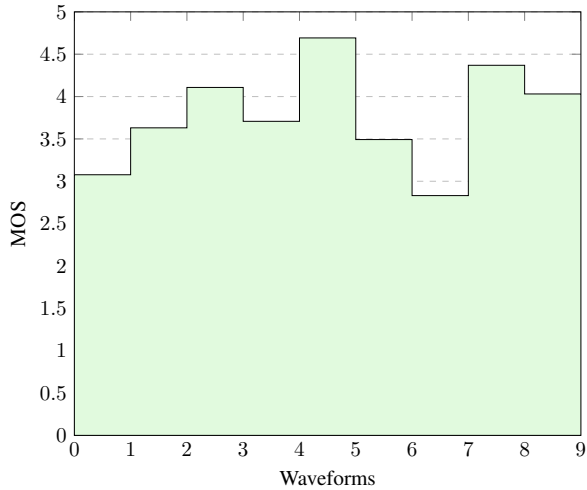
\begin{figure}[h]
    \centering
    
    \begin{tikzpicture}[thick, scale = .8]
\begin{axis} [
    xlabel = Waveforms,
    ylabel = MOS,
    xmin=0, xmax=9,
    ymin=0, ymax=5,
    xtick={0,1,2,3,4,5,6,7,8,9},
    ytick={0,0.5,1,1.5,2,2.5,3,3.5,4,4.5,5},
    ymajorgrids=true,
    grid style=dashed,
]
\addplot[
    const plot,
    fill=lightgreen,
    draw=black
]
coordinates {
(0, 3.076923076923077)
(1, 3.6307692307692307)
(2, 4.107692307692307)
(3, 3.707692307692308)
(4, 4.6923076923076925)
(5, 3.4923076923076923)
(6, 2.830769230769231)
(7, 4.369230769230769)
(8, 4.030769230769231)
(9, 3.953846153846154)
}
\closedcycle;
\end{axis}
\end{tikzpicture}
\label{stimuli_mos}
    \caption{MOS of 10 synthesized wave}
    \label{fig:mos_stimuli}
\end{figure}

MOS measured the naturalness of a TTS system. Five labels are given to some users for scoring the system. These five labels are Excellent, Good, Fair, Poor, and Bad. After listening to the waveforms, the individual user chooses a label to rate each of the synthesized waveforms.  
\begin{table}[h]
    \centering
    \begin{tabular}{|l|l|}
        \hline
        \textbf{Label} & \textbf{Rating}\\
        \hline
        Excellent & 5\\
        \hline
        Good & 4\\
        \hline
        Fair & 3\\
        \hline
        Poor & 2\\
        \hline
        Bad & 1\\
        \hline
    \end{tabular}
    \caption{Label vs Rating}
    \label{map}
\end{table}
These five labels are mapped with some pre-defined numbers. Table \ref{map} shows the mapping between these five labels and the numbers. After that, the arithmetic mean of those numbers has been calculated using the equation \ref{eq:MOS} which we called MOS. 
\begin{equation}
\label{eq:MOS}
MOS = \frac{1}{N} \sum_{n=1}^{N} R_n
\end{equation}
Individual rating is an integer number but the final result can be a real value. For rating our system, 65 native Bengali speakers from different backgrounds participated. We provided 10 waveforms to them generated by our TTS system. Those 10 waveforms are available at this Github\footnote{https://sustbn.github.io/} link. They gave a label-based rating to those 10 waveforms based on naturalness. So, we received 65 ratings for each of the 10 waveforms and calculated an average rating of the individual 10 waveforms. Finally, we calculated the arithmetic mean of those 10 individual ratings using the formula \ref{eq:MOS} and got 3.79 as Mean Opinion Score (MOS). Figure \ref{fig:mos_stimuli} shows the individual MOS of 10 waveforms.

\subsection{Discussion and Analysis}
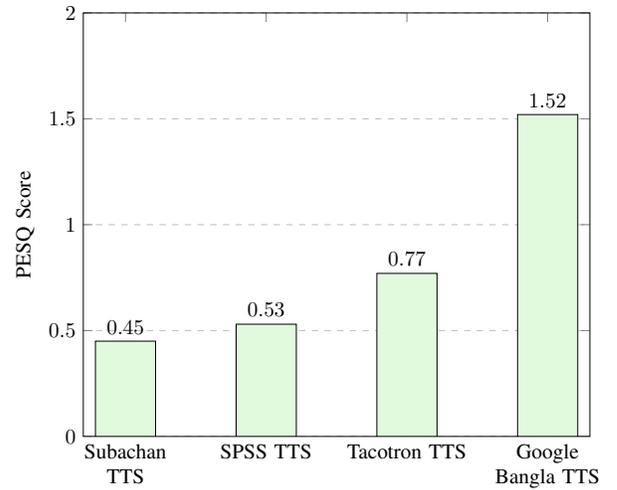
\begin{figure}[H]
    \centering
    \begin{tikzpicture}[thick, scale = .8]
    \begin{axis}[
        symbolic x coords={Subachan TTS, SPSS TTS, Tacotron TTS, Google Bangla TTS},
        ylabel=PESQ Score,
        bar width=1cm,
        ymin=0,
        ymax=2,
        nodes near coords,
        ytick={0,0.5,...,2},
        ymajorgrids=true,
        grid style=dashed,
        xtick=data,
        x tick label
        style={rotate=0,anchor=north,align=center,text width=2cm},
        ]
        \addplot[ybar, fill=lightgreen] coordinates {
                (Subachan TTS,   .45)
                (SPSS TTS,  .53)
                (Tacotron TTS,   .77)
                (Google Bangla TTS, 1.52)
        };

\end{axis}
\end{tikzpicture}
\caption{PESQ Scores of Different TTS Systems}
\label{fig:pesq_com}
\end{figure}

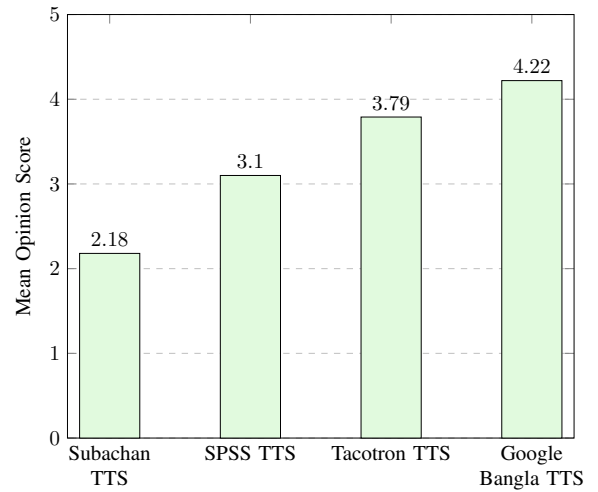
\begin{figure}[h]
    \centering
    \begin{tikzpicture}[thick, scale = .8]
\begin{axis}[
        symbolic x coords={Subachan TTS,SPSS TTS,Tacotron TTS,Google Bangla TTS},
        ylabel=Mean Opinion Score,
        bar width=1cm,
        ymin=0,
        ymax=5,
        nodes near coords,
        ytick={0,1,...,5},
        ymajorgrids=true,
        grid style=dashed,
        xtick=data,
        x tick label
        style={rotate=0,anchor=north,align=center,text width=2cm},
        ]
        \addplot[ybar, fill=lightgreen] coordinates {
                (Subachan TTS,   2.18)
                (SPSS TTS,  3.1)
                (Tacotron TTS,   3.79)
                (Google Bangla TTS, 4.22)
        };
\end{axis}
\end{tikzpicture}
    \caption{MOS of Different TTS Systems}
    \label{fig:mos_com}
\end{figure}

We are now going to show the graphical comparisons of various existing Bangla TTS systems. We calculated PESQ and MOS for our Tacotron TTS as well as for Google Bangla TTS \cite{gutkin2016tts}, SPSS TTS \cite{raju2019bangla}, and Subachan TTS \cite{naser2010implementation}. The bar chart of figure \ref{fig:pesq_com} shows the comparison of PESQ scores of different TTS systems. Subachan TTS and SPSS TTS have obtained 0.45 and 0.53 PESQ scores respectively. Our current Tacotron TTS has achieved a 0.77 PESQ score which is better than noncommercial SPSS TTS and Subachan TTS. Here we can observe a gradual advancement of our system. Another comparison of various TTS systems is shown in figure \ref{fig:mos_com} concerning Mean Opinion Score (MOS). We have achieved 3.79 MOS which is better than the previously implemented systems like Subachan TTS and SPSS TTS. There are no significant works on Bangla TTS using deep learning. A very recent work Byakto Speech \cite{nazi2021byakto} has obtained a 3.23 MOS score. In some cases, it is even comparable with commercially implemented Google Bangla TTS.

\section{Conclusion}
Tacotron TTS or end-to-end Bangla TTS is the current state-of-the-art TTS system. It overcomes the drawbacks of the traditional Deep Neural Network (DNN) based TTS system which needs more building blocks, hand-engineered features, etc. Based on naturalness, achieving a 3.79 score as MOS is very significant. This score outperforms all existing non-commercial Bangla TTS systems. Moreover, based on the objective evaluation we got a 0.77 PESQ score which also correlates with the MOS. We used the Griffin-lim spectrogram inversion algorithm which is till now the fastest one. Though text normalization has been done before feeding the data into the network, modern research is proposing the automation of text normalization.

\bibliographystyle{IEEEtran}
\bibliography{ref}

\end{document}